Physics Magic
Nathaniel Lasry and Pierre-Osias Christin

The purpose of this paper is to show the magic of physics by showing the physics of magic. What usually makes magic tricks interesting is that something unexpected occurs. Similarly, demonstrations are interesting inasmuch as they produce something unexpected. Since expectations are linked to preconceptions, a demonstration making use of a flaw in a preconception will result in something unexpected. Given the numerous misconceptions in physics, many demonstrations can be dressed up as magic tricks.

The first objective of this paper is to share with other physics teachers the excitement of creating and using magical classroom demonstrations. The second objective is to provide interested instructors with practical means to convert a classical demonstration into a magic trick. To illustrate the procedure, two classical demonstrations will be re-presented as the magic tricks we have presented in our courses. The final goal is to use current ideas in educational psychology to explain why using magic has worked so well in our courses in providing students with a new impetus to learn physics. This description is not meant to be formal, but proposes a theoretical model that fits our classroom observations.

*Addressing Misconceptions*

Students entering a high school or college physics classrooms possess large bodies of knowledge. However, this previous knowledge is often incompatible -and may even interfere- with correct physics concepts[2]. Many of the prevalent misconceptions that students have with respect to classical physics have been identified[1] and conceptual learning has been shown to be facilitated by explicitly addressing misconceptions[3]. The suggestion in this paper is to design situations where the students' misconceptions are used to create an expectation which is not met. Essentially, the instructor is saying: "*here's what you expect; now here's what happens...*". Why should a departure from expectation facilitate learning?

There are many possible responses to such a question. The reader may identify with personal experiences where, faced with something puzzling (i.e. unexpected), an inquiry into the underlying process or phenomenon was initiated. As a personal anecdote: one of us (NL) recalls being about 8 years old and seeing city workers fixing a sidewalk approximately 100m down the street where he lived. One worker would repeatedly raise a sledgehammer above his head and then pound the sidewalk. Yet, the noise of the pounding lagged behind the pounding itself. This was like one of those annoying off-sync movies (there were a few on TV about 25 years ago…). But it wasn't a movie! Why was the worker off-sync? It was this unexpected observation that pushed the 8 year old to inquire about the physical phenomenon and begin to conceptualize the significant difference between the speed of sound and the speed of light.

Although personal experiences may yield the most convincing rationales, one of the greatest lessons of physics education research is that rationales based on personal experience should be taken cautiously. After all, student misconceptions are mostly derived from personal experiences. Therefore, the question essentially remains: Why should a departure from expectation facilitate learning?

Since physics is quite counter-intuitive to students, one of the major goals in physics education is to enable learning through conceptual change. Cognitive conflict, the reaction to a situation that cannot be accounted for within a set of preconceptions, is a process that triggers conceptual change[4]. Yet, magic demonstrations are predicated on the necessity to causes something unexpected to occur and thus implicitly use cognitive conflict. The physics education literature suggests that the cornerstone of effective instruction is to make students' preconceptions explicit and then address them head-on[3,5]. In a magic demonstration, a preconception is most often made explicit to build up the tension and the drama of the finale: the unexpected observation. This observation becomes the source of the cognitive conflict and can be used as a starting point into an inquiry cycle whereby learners will attempt to accommodate and assimilate a newly constructed concept.

One may also contend that a magic demonstration is a source of interest and fun for students. Thus, beyond cognitive effects, magic arouses emotions that are not among the palette of emotions normally found in higher education. Interestingly, from a neuro-cognitive perspective, new (declarative) knowledge is processed through a brain structure called the Hippocampus[6]. Yet, the Hippocampus is also a central part of the limbic system: the mind's emotional processing unit[7]. Since new knowledge is encoded by the structure that processes emotions, a magic demonstration may optimize the arousal of cognitive and affective processing and thus enhance the learning experience. Since the mind can be seen as the interplay of cognition, affect (emotions) and conation (motivation)[8], magic can also be seen as a symbiotic process that addresses these three dimensions: cognition through cognitive conflict ; emotions through the reactions to the tricks and motivation as the resulting desire to inquire into the functioning of the tricks.

*Why modify demonstrations?*

Demonstrations are often used in classrooms as great tools to liven up a lecture. Students seem to perk up and pay closer attention. Yet, recent findings suggest that students passively observing a demonstration learn no more than those who have not seen the demonstration[9]. For demonstrations to be effective, students must be actively engaged by for instance, making a prediction as to what will happen. Making a prediction engages students actively: they must make their preconception explicit to formulate a hypothesis and usually develop a vested interest in the outcome (*"will I be right?"*). When the outcome differs from their prediction, a cognitive conflict may be experienced and the likelihood of learning is increased.

Magic demonstrations also seem to actively engage students by making prevalent pre-conceptions explicit to create an unexpected observation. If the tension and drama are properly built into the trick, cognitive conflict is quite likely to occur. Furthermore, it is useful to have students learn these tricks and, although students might do this for the social benefits (such as performing party tricks), hands-on experience with these demonstrations is likely to contribute to conceptual change.

*How to make a Demo Magical*

Three demonstrations are presented below. The most classical demonstration, the bed of nails, will be presented first, followed by the double conic roller (center of mass) and the 'magical beans' (buoyancy/density). Although, these demonstrations may be familiar to seasoned teachers, the purpose here is to show a different mode of presentation. Thus, the main purpose is not to sell the salad but the dressing.

*The bed of nails*

Traditionally, this demonstration is used to illustrate the concept of pressure as the ratio of force over surface area. A formal or conceptual presentation of the concept is usually followed by the demonstration. The first suggestion here is to reverse this sequence.

Presenting the bed of nail as a magic demonstration may be performed by suggesting that the instructor has discovered the path to occult powers. After tying a turban on one's head to tune into the students' mental 'vibes' and allow levitation to occur, a brief meditation moment is taken. The instructor, displaying much hesitation, states his/her apprehension of doing this magical feat and decides to take a bite out of an apple to relax. The apple is clumsily dropped onto the bed of nails and retrieved perforated, showing the sharpness of the nails. The dropped apple makes explicit the preconception that anything in contact with the nails will be perforated. Students are then asked to focus their "mental energies" and "send positive vibes". While chanting a mantra, the instructor, encouraged by the students, resolves (not without hesitation) to lie down on the bed. Students expecting the worse (or best, depending on the student and the outcome…) observe the instructor lying down onto the bed of nails and may then be asked to explain why the instructor was not transformed into a sifter. The inquiry cycle may then begin.

There is an interesting complement to this demonstration. Once the concept of pressure seems to have been properly understood, students may be asked to predict whether the instructor could walk on the bed of nails without shoes. The instructor positively acknowledges responses contradicting the possibility of walking on nails and proceeds to explain that it would be impossible since the instructors' weight could not possibly rest on such a small surface as a foot without being perforated (at least not with regular surface density of nails). During this explanation, the instructor removes his/her shoes (but not the socks) and proceeds to walk on the bed of nails, emphasizing the impossibility of the feat at every step. When students ask why it is that the instructor is able to walk on the nails, the instructor steps down from the bed of nails, removes the socks and reveals 2 cardboard insoles hidden inside the socks and ends with a suggestion that critical thinking requires one to go beyond appearances.

*Deconstructing the demo*

This demonstration is an ideal magic demonstration since it is traditionally performed by magicians and fakirs. The effectiveness of the demonstration lies in the preconception that anything in contact with nails will be perforated. Dropping the apple contributes to making that expectation even more explicit. Furthermore, hesitation towards performing the trick also adds to the tension as a number of students may urge the instructor not to proceed (or to go ahead depending on the student…). Thus, the primary aim is to trigger a cognitive conflict with the preconception and allow students to conceptualize pressure as the ratio of force over a surface area.

Once the students seem to have understood the concept, the instructor can then present them with the second part, an application of the concept to a slightly different context: can one walk on the nails (i.e. same weight but much smaller surface area)? Once again, this portion is set up to maximize cognitive conflict as the students having acknowledged that the pressure is inversely proportional to the surface area are conscious of the difficulty of walking on a bed of nails. The conflict is maximized when the discourse of the instructor is in direct conflict with the students' observation; until the cardboard insoles are revealed and the impossibility of walking on nails is confirmed.

*The double conic roller*

Among great 'center of mass' demonstrations available from most lab equipment providers is the double conical roller placed on an *inclined wedge*.

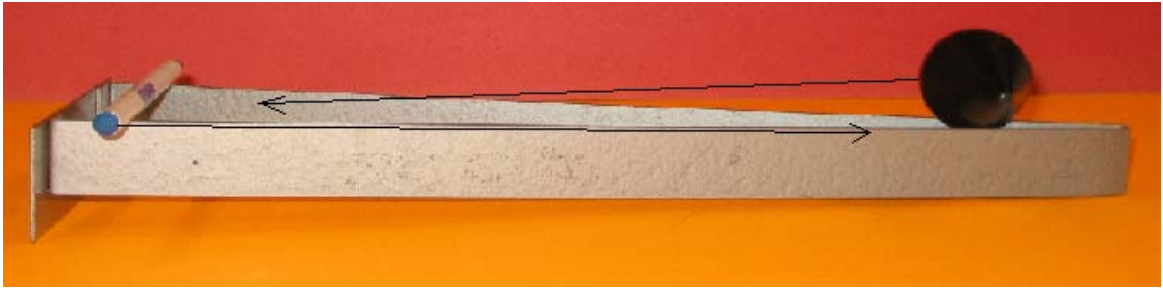

If one places a pen on the elevated side of the wedge, it will roll down the incline towards the narrower part of the wedge. However, when the double cone is placed on the "bottom" of the incline, it will roll "up" the incline. This is due to the fact that the center of mass of the double cone is propped up where the wedge is narrower although this happens to be the bottom of the incline. The roller sinks into the wedge as it gets wider towards the 'top' of the incline. This gives the impression that the double-cone rolls up the incline when in fact it is rolling down into the wedge. So the pen seems to move down the incline while the double cone seems to move up the incline. Usually, this demonstration is presented and center of mass dynamics are discussed.

To present this demo as a magic trick, one shows the inclined wedge and wood double cone. It is useful to hand these out to 1 or 2 students for inspection so they can make sure that there are no "gimmicks" (students usually look for magnets). The instructor then shows that one part of the incline is elevated by letting a pen roll down the incline.

With a deep breath and tremendous concentration, the instructor states that, with the mind's power, telekinesis will be performed, and the double cone will be… dragged UP the incline, 'against' gravity (drum roll…). The double cone is then released from the bottom of the incline and rolls towards the top of the incline. This portion constitutes the unexpected, the "*here's what actually happens*" part of the demonstration.

As the double cone proceeds up the incline, the instructor's hands precede it and seem to be magically pulling it upwards. When the double cone reaches the top, the instructor quickly picks up the double cone (since the "trick" would be spoiled if the cone just sat there for a lengthy period of time). A long sigh is released indicating the tremendous mental effort that was required.

Usually, students will ask this trick to be performed again. It is useful to repeat the trick once or twice, so that students may focus on different aspects of what they believe to be happening. Students usually volunteer explanations and it is useful to ask the class to debunk the trick through some structured inquiry process.

*Deconstructing the demo*

This first part of this trick is making the preconception explicit: the "***here's what you expect***" portion of the demo. The importance of this part cannot be overemphasised since without clear expectation, a strong cognitive conflict cannot arise. In this demonstration, the pen is rolled down the incline to show that one side is 'higher' than the other. Thus, the demonstration shows the flaw in the pre-conception that *"what constitutes "high" and "low" in determining the direction in which an object will fall is absolute: if it is High for one object it must be High for all objects"*. This demonstration shows that holding this belief generates something unexpected since the direction in which an object will fall is not dictated by the surface of a support but by the path of the center of mass of the object. Thus, two objects on an identical support may have different "highs" and "lows" and may therefore fall in different directions. This observation should trigger cognitive conflict and plant the seeds of inquiry.

***Conclusion***

Presenting demonstrations as magic tricks is effective inasmuch as the pre-conceptions are made explicit and are shown to be ineffective in explaining some observation. In a sense, these demos say "Here's what you expect. With a wave of the hands and an Abracadabra… it didn't happen!". The process of turning a demonstration into a magic trick can be applied to a large number of demos. The main requirement is that a preconception can be determined and made explicit and that the observation will be unexpected. Demonstrations can also involve deception to make the observation even more unexpected. For instance, we have modified the classical Bernoulli's principle demo where a ping-pong ball floats above a blow dryer by hiding the blow dryer below a table. With magic ambiance music sufficiently loud to cover the blow dryer's noise, the instructor can then release a ping pong ball in the air and make it levitate. Class time can then be devoted to "debunking" the trick and an inquiry process can de initiated. From a cognitive perspective, creating such unexpected situations sets the stage for the cognitive conflict which should facilitate conceptual change. That is, the cognitive conflict should cause students to reassess their existing model and eventually change their conception to accommodate for the unexplained observation. However good this cognitive argument, you may be inclined to respond to the affective argument (that also provided the impetus for writing this paper): using magic demos in class is simply too much fun!